\documentclass[pra,aps,superscriptaddress,notitlepage,longbibliography,nofootinbib,twocolumn]{revtex4-2}
\usepackage[margin=1.54cm]{geometry}
\usepackage{wrapfig}
\usepackage{graphicx}
\usepackage{mathtools}
\usepackage{bm}
\usepackage{amssymb}
\usepackage{epstopdf}
\usepackage{siunitx}
\usepackage{xcolor}
\usepackage[ngerman,english]{babel}
\usepackage[T1]{fontenc}
\usepackage{amsmath}
\usepackage{bbold}
\usepackage{hyperref}
\usepackage{comment}
\usepackage[normalem]{ulem}
\usepackage{CJKutf8}
\usepackage{braket}
\usepackage{tikz}
\usetikzlibrary{positioning}

\graphicspath{{./figs/}{./figures/}}

\DeclareUnicodeCharacter{2009}{ }

\usepackage{hyperref}
 \hypersetup{
     colorlinks = true,
     linkcolor  = black,
     filecolor  = blue,
     citecolor  = magenta,      
     urlcolor   = black,
     }

\usepackage{braket}

\newcommand{\sigp}{\sigma^+}
\newcommand{\sigm}{\sigma^-}

\definecolor{orange}{RGB}{200,100,0}

\usepackage[markup=underlined]{changes}

\usepackage{letltxmacro}
\LetLtxMacro{\ORIGselectlanguage}{\selectlanguage}
\makeatletter
\DeclareRobustCommand{\selectlanguage}[1]{%
  \@ifundefined{alias@\string#1}
    {\ORIGselectlanguage{#1}}
    {\begingroup\edef\x{\endgroup
       \noexpand\ORIGselectlanguage{\@nameuse{alias@#1}}}\x}%
}
\newcommand{\definelanguagealias}[2]{%
  \@namedef{alias@#1}{#2}%
}

\makeatother

\definelanguagealias{en}{english}
\definelanguagealias{EN}{english}
\definelanguagealias{eng}{english}
\definelanguagealias{de}{ngerman}

\begin{document}

\title{Reconciling quantum and classical spectral theories of ultrastrong coupling:
\\
 Role of cavity bath coupling and gauge corrections}

\author{Stephen Hughes}
\email{shughes@queensu.ca}
\affiliation{Department of Physics, Engineering Physics and Astronomy, Queen's University, Kingston, ON K7L 3N6, Canada}
\author{Chris Gustin}
\email{cgustin@stanford.edu}
\affiliation{Edward L.\ Ginzton Laboratory, Stanford University, Stanford, California 94305, USA}
\email{cgustin@stanford.edu}
\author{Franco Nori}
\email{fnori@riken.jp}
 \affiliation{Theoretical Quantum Physics Laboratory, Cluster for Pioneering Research, RIKEN, Wako-shi, Saitama 351-0198, Japan}
  \affiliation{Center for Quantum Computing, RIKEN, RIKEN, Wako-shi, Saitama 351-0198, Japan}
\affiliation{Physics Department, The University of Michigan, Ann Arbor, Michigan 48109-1040, USA}

\date{\today}

\begin{abstract}
Focusing on the widely adopted 
Hopfield model with cavity dissipation,
   we show how the linear spectrum of 
  an ultrastrongly  coupled cavity and a dipole can be described either
   classically or quantum mechanically,
  but only when the quantum model includes (i) corrections to maintain gauge invariance, and  (ii) a specific type of cavity bath coupling.
   We  also show the impact of this bath model on the quantum Rabi model, which has no classical analogue in ultrastrong coupling.  
\end{abstract}

\maketitle

{\it Introduction.}---Strong coupling
between a single cavity mode
and a dipole or two-level system (TLS)~\cite{PhysRevLett.53.1732,PhysRevLett.93.233603,PhysRevB.60.13276} can be well explained quantum mechanically or classically (or semiclassically) \cite{PhysRevLett.64.2499,RevModPhys.71.1591,PhysRevB.70.195313}. 
The  characteristic signature of strong coupling is a splitting in the emitted spectrum 
by $2g$, where $g$ is the  dipole-cavity coupling rate,
which exceeds any losses in the system, e.g.,
$g^2>\kappa^2/16$~\cite{PhysRevB.60.13276} (or more strictly $g^2>\kappa^2/8$~\cite{PhysRevLett.111.053901}), with $\kappa$  the cavity decay rate.
Quantum mechanically, this is referred to as
{\it vacuum} Rabi splitting, or in classical physics as normal-mode splitting. 

Strong coupling  has been observed
in 
atoms~\cite{PhysRevLett.93.233603},
molecules~\cite{herrera_cavity-controlled_2016,flick_atoms_2017},
quantum dots~\cite{reithmaier_strong_2004,yoshie_vacuum_2004}, and circuit QED~\cite{gu_microwave_2017,cao2010dynamics}, and is often considered a prerequisite for exploring {\it unique quantum effects} when one moves beyond a weak
excitation approximation or 
linear response~\cite{schuster_nonlinear_2008,bose_all-optical_2014}.
In a quantum description of  
cavity-TLS coupling, 
multi-photon effects manifest in an {\it anharmonic} response~\cite{Fink2008,
Bishop2008,Illes2015,hamsen_two-photon_2017}, which is not captured by the physics of two coupled classical harmonic oscillators (HOs). Nevertheless,
a classical description of the emitted spectrum
under weak excitation is an adequate
description of the system, and one recovers a perfect quantum to classical correspondence of the light-matter system, albeit with a different interpretation.
Indeed, 
the quantum interpretation of spontaneous emission  can be described in terms of radiation reaction, vacuum fluctuations, or a mixture of both these effects~\cite{PhysRevA.11.814,Milonni1976}.

Quantum and classical descriptions of certain light-matter coupling have lead to 
interesting interpretations and 
insights,
including the difference between 
quantum and classical oscillations in 
phase qubits
\cite{PhysRevB.78.054512},
classical pseudo-Rabi oscillations in 
flux 
qubits 
\cite{PhysRevB.78.054512},
and vacuum Rabi splitting as a
manifestation of linear-dispersion theory
\cite{PhysRevLett.64.2499}.
%

Beyond strong coupling, recent interest in
cavity-QED has turned to ultrastrong coupling (USC)~\cite{PhysRevA.74.033811,anappara_signatures_2009,beaudoin_dissipation_2011,PhysRevA.80.033846,de_bernardis_breakdown_2018,frisk_kockum_ultrastrong_2019,forn-diaz_ultrastrong_2019, LeBoit2020}, where one cannot invoke a rotating wave approximation (RWA), typically
when $g \geq 0.1\omega_0$, where $\omega_0$ is the dipole 
resonance frequency. 
The regime of USC presents some fascinating 
uniquely quantum concepts such as {\it virtual photons in the ground state}~\cite{PhysRevLett.98.103602,forn-diaz_ultrastrong_2019,LeBoit2020,frisk_kockum_ultrastrong_2019}. 
Squeezed vacuum states, also with no classical analogue, occur in both bosonic and TLS emitter systems in 
the USC, which are described using the Hopfield model and the quantum Rabi model (QRM), respectively~\cite{forn-diaz_ultrastrong_2019}. 
Exciton and many-emitter Dicke systems also take on the form of the Hopfield model, such as cavity coupling to
2D electron systems including
Landau levels in 
THz cavities~\cite{Zhang2016}.
In the USC regime, these systems exhibit spectral signatures that reflect the nature of the quantum Hopfield model.
On the other hand, 
the classical theory of coupled oscillators (including collectively coupled TLSs in the dilute thermodynamic limit) does not require a RWA, and  it might be expected that the USC regime should also have a quantum to classical correspondence. 

 In probing the Hopfield regime, the optical spectrum is typically measured, and most features are argued as being quantum mechanical in nature, e.g., stemming from diamagnetic coupling terms. Yet, 
 the spectral locations of the upper and lower polaritons can be matched with coupled classical HO theory \cite{PhysRevA.74.033811,Rajabali2021,Garziano2020Aug}. In the presence of dissipation, 
  however,
a quantum-classical correspondence 
is not known.
Although there exists some quantum Langevin approaches for simplified geometries
\cite{PhysRevA.74.033811}, 
these are not suitable for calculating higher-order quantum correlation functions and arbitrary geometries.
Furthermore, 
the physics of a cavity-coupled TLS is often said to reduce to cavity-HO physics if one neglects saturation effects~\cite{Bouchet2019,Jrgensen2022}. However, this is not true in the USC regime, even with linear response, 
since multiple photon states
emerge already in excited states,
and saturation effects are 
unavoidable in the USC regime due to the virtual excitation of the TLS, including the ground state. 

The USC regime presents additional challenges for quantum field models, including:
(i) 
gauge corrections because of a truncated Hilbert space~\cite{de_bernardis_breakdown_2018,di_stefano_resolution_2019,PhysRevA.107.013722}, and (ii) the specific form of the system-bath interaction for the cavity mode 
matters~\cite{Bamba2014Feb,Salmon2022}. Since a classically-coupled HO model (expressed entirely in terms of  classical electromagnetic fields) has no issues with gauge invariance, it is
essential to seek out if and when such a quantum-classical correspondence can be made. 
This is not just motivated by fundamental physics reasons, 
but  is  practically important since many of the emerging USC experiments require some sort of modelling with classical Maxwell solvers 
\cite{Baranov2020,Rajabali2021}.


In this work, we 
first show that, for a lossless system, the spectral poles
(resonances) 
of the Hopfield model precisely overlap a classical 
HO solution, and these deviate from the QRM as soon as one enters the USC regime. We then 
introduce a 
spectral theory
of the dissipative Hopfield model,
using a gauge-invariant master equation theory expressed in the multi-polar gauge, and show
how it is possible 
to identify 
a specific form of the system-bath coupling 
that 
matches the classical solution.

\begin{figure}[t]
    \centering
\includegraphics[width=0.85\linewidth]{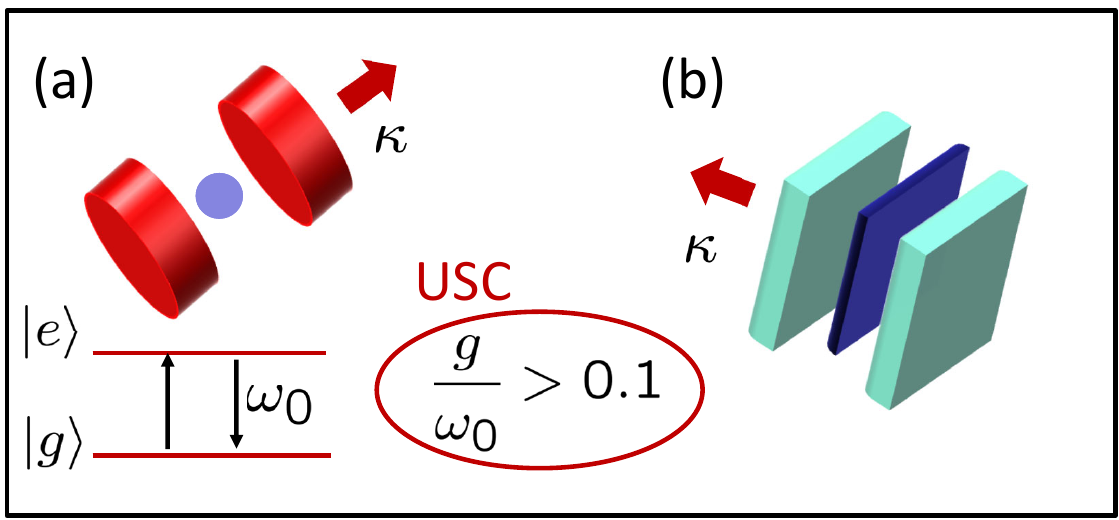}
    \caption{Schematic of 
    two systems in dissipative cavity-QED that can realize USC, including (a) an
    atom inside a cavity, which has a decay rate $\kappa$, and  (b) a planar cavity coupled to a collective emitter system. The emitters can be treated as a bosonic  (Hopfield model) or as a
    TLS (QRM).
    }
    \label{fig:schematic}
\end{figure}

We choose
a common and established
classical theory,
based on a normal-mode expansion of the 
cavity Green function with 
phenomenological decay; a more rigorous approach could 
use
quasinormal modes~\cite{
PhysRevLett.122.213901,PhysRevResearch.2.033456}.
Finally, we study the impact 
of this model on 
both the dissipative Hopfield model and the dissipative QRM, 
 and show the striking differences between these two models for different $\eta \equiv g/\omega_c$. 
  Usually
$\eta>0.1$ is the criterion for the USC regime. 
 A simple schematic 
of lossy cavity-QED systems are shown in Fig.~\ref{fig:schematic}.


{\it Theory.---}We first consider the interaction between 
a bosonic cavity mode, with creation (annihilation)
operator $a^\dagger, a$,
and a bosonic dipole,
with creation (annihilation)
operator $b^\dagger, b$.
Neglecting bath losses for now,
in the multi-polar gauge,
and with the dipole approximation, 
the Hopfield model can be written as 
($\hbar=1$)
\begin{equation}\label{eqn:Hop1}
    H_{\rm Hop} = \omega_c a^\dagger a + \omega_0 b^\dagger b + ig(a^{\dagger}-a)(b+b^\dagger)
    + D (b+b^\dagger)^2,
\end{equation}
where $\omega_c$ ($\omega_0$) is the cavity (atom) transition frequency,
and
$D=\eta g$ is the diamagnetic amplitude~\cite{2002.02139}. {Note that a naive truncation of the multipolar gauge Hamiltonian would render this Hopfield $D$ term infinite, and one must account for the gauge invariance of the truncated single-mode model  to obtain physical and correct results~\cite{PhysRevA.107.013722,Taylor2022Mar}.}

Using  a Bogoliubov transformation,
we rewrite this Hamiltonian as
$ H_{\rm Hop} = \omega_c a^\dagger a + \tilde \omega_0 \tilde b^\dagger \tilde b + i \tilde g(a^{\dagger}-a)(\tilde b+\tilde b^\dagger) + D$,
where $\tilde \omega_0 =
\omega_0(1+4 \eta^2)^{0.5}$
and
$\tilde g^2 =g^2/(1+4 \eta^2)^{0.5}$.
Diagonalization yields two polariton poles~\cite{Emary2003Jan}:
$\omega^2_\pm =
\frac{1}{2}
\left[\tilde \omega_0^2 + \omega_c^2
\pm \sqrt{(\tilde \omega_0^2-\omega_c^2)^2+16\tilde g^2 \tilde\omega_0\omega_c}\right]$.
Assuming on resonance conditions
($\omega_0=\omega_c$), then
\begin{equation}\label{eq:polariton_poles}
    \omega_\pm = \omega_0\sqrt{1 + 2\eta^2 \pm 2\eta(1+\eta^2)^{1/2}}.
\end{equation}
To lowest order
in the counter rotating-wave effects, i.e., to order $\eta^2$
(Bloch-Siegert regime),
we obtain
$\omega_\pm|_{\rm BS} = \omega_0(1 + \eta^2/2)
\pm g$.
If one  neglects the
diamagnetic term, then
$
\omega_\pm^0 =\omega_0(1\pm 2\eta)^{0.5}$,
which is  problematic for 
$\eta\geq 0.5$. 

We can also compare this solution to
the QRM, with
\begin{equation}\label{eqn:QR}
    H_{\rm QRM} = \omega_c a^\dagger a + \omega_0 \sigp \sigm + ig(a^{\dagger}-a)(\sigp+\sigm),
\end{equation}
where $\sigp$ ($\sigm$) is the creation (annihilation) operator for the TLS, 
which has important saturation 
effects.
In this case, we have no diamagnetic term to consider as the relevant TLS term has no effect on energy differences with the TLS operators.
For the QRM, we have an infinite
set of anharmonic eigenenergies, which modifies the 
Jaynes-Cummings ladder states because of counter-rotating wave terms. Considering again a Bloch-Siegert regime (order $\eta^2$ coupling)~\cite{LeBoit2020,Salmon2022}, we obtain the poles
of the lowest order polaritons
$\omega_\pm |_{\rm BS}^{\rm QRM} 
= \omega_0 \pm g(1+\eta^2/4)^{0.5}
\approx  \omega_0 \pm g$.
These QRM  pole
resonances differ from the
Hopfield model in the Bloch-Siegert regime, even with linear response.

Linear spectral shifts beyond those in the
Jaynes-Cummings model are 
often termed
vacuum Bloch-Siegert
shifts~\cite{Li2018}, but below we  quantify why there is nothing uniquely quantum  about such resonance shifts in a Hopfield model.
This is in contrast to the QRM, which becomes uniquely quantum in nature 
in the USC regime.


In classical electromagnetism, 
the bare polarizability volume of an oscillator is
$\alpha(\omega) = {A_0 \omega_0}/
{(\omega_0^2-\omega^2)}$,
with 
$A_0 = {2d^2}/{\epsilon_0}$
and $d$ the dipole moment.
Considering the 
emitter position
${\bf r}_0$,
 the photonic Green function, under a single mode expansion (and assuming scalar fields) is~\cite{SI}
\begin{equation}
    G_c({\bf r}_0,{\bf r}_0,\omega) = \frac{A_c\omega^2}{\omega_c^2-\omega^2},
\end{equation}
where  $A_c=1/V_{\rm eff}\epsilon_b$, with $V_{\rm eff}$ the effective mode volume and 
$\epsilon_b$ the background dielectric constant.
Embedding the HO dipole in the cavity, we 
obtain the exact polarizability:
\begin{equation}
    \alpha(\omega) 
    = \frac{A_0\omega_0 }{\omega_0^2-\omega^2-
    (\omega_0/\omega_c)4 g^2\omega^2/(\omega_c^2-\omega^2)},
\end{equation}
where $4g^2 = d^2\omega_c/(2\epsilon_0V_{\rm eff} \epsilon_b)$.
Considering the on-resonance case
again, 
 the classical poles are 
 $\omega_{\pm}^G = \omega_{\pm}$ [as in Eq.~\eqref{eq:polariton_poles}],
so {\it it is identical} to the solution of the  lossless Hopfield model. This is consistent with experimental results on ultrastrongly-coupled molecular vibrational dipoles in IR cavities, where the same classical-quantum correspondence was observed in the oscillator frequencies~\cite{George2016Oct}.
In a quantum picture, the blueshift is caused by the 
$P^2$ term (or $A^2$ term in the Coulomb gauge). However, in a classical picture, this blueshift is  caused from the poles of the cavity-renormalized polarizability; thus, there is nothing uniquely quantum about this spectral blueshift.
The blueshift is caused by counter rotating-wave effects, in both models.

This correspondence with the poles
of the quantum Hopfield model
and classical electromagnetism is  partly
known~\cite{Rajabali2021,Garziano2020Aug}, yet sometimes not recovered in quantum models.
Moreover, in a linear optical material, the classical Green function of the hybrid system must have poles in the upper complex plane~\cite{Bamba2013Jul}, even with 
 linear gain~\cite{PhysRevLett.127.013602}.
 The classical pole correspondence does not mean that there are no unique quantum
effects in the Hopfield model, since the ground and excited states are squeezed states~\cite{PhysRevLett.98.103602,Makihara2021}.
Indeed, for $\omega_c=\omega_0$, 
the quantum ground state, $\ket{0_+0_-}$
has energy:
$\omega_{0,0}{=}(\omega_+{+}\omega_-)/2 {-} \omega_0
=\omega_0(1+\eta^2)^{0.5} {-} \omega_0>0$.
The ground state contains virtual photons 
and is an entangled state~\cite{2304.00680}. 
Despite this, there appears to be no unique quantum effects that affect the polariton eigenfrequencies.

What is less well known is to what degree the predicted optical spectra agree (or not) between the quantum and classical coupled mode theories in the USC regime, and how to describe such a regime with open-system master equations.
Since all cavity-QED systems have dissipation and input-output channels, it is essential to model them as open quantum systems. 
Within the RWA,
the vacuum Rabi doublets are well described classically or quantum mechanically \cite{PhysRevLett.64.2499}, for both boson and TLS emitters.
In the USC regime, things are much more subtle, and
the quantum models have  technical problems related to how to properly include dissipation as well as gauge correction terms (caused by material and cavity mode truncation). 


In a classical light-matter  model,
we can include a heuristic cavity decay rate,
$\kappa$, in the cavity Green function, and
 derive the
the cavity-emitted 
spectrum as 
\begin{equation}
     S^{\rm Class} =  F({\bf R})   \left|
 \frac{E_0\,g^2\omega^2}{(\omega^2-\omega_c^2 - i\omega\kappa)(\omega^2-\omega_0^2)- 4g^2 \omega^2} \right
 |^2, 
 \label{eq:class1}
 \end{equation}
 where $E_0$ is the 
excitation  field strength 
and
$F({\bf R})$ is  a geometric factor.
The solution is
non-Markovian, causal, and contains no RWA~\cite{SI}. 
Also, this phenomenological 
approach to 
dissipation  ensures a symmetric spectrum outside of the USC regime for a resonant cavity and TLS.

In a quantum picture,
to include cavity dissipation, we use 
an open-system approach~\cite{carmichael_statistical_2013,Salmon2022},
at the level of a generalized master equation (GME)~\cite{settineri_dissipation_2018,Salmon2022,PhysRevResearch.4.023048,PhysRevResearch.5.033002},
\begin{equation}
\label{eq:GME1}
\frac{\rm{d}}{\rm{dt}}\rho = -\frac{i}{\hbar}[H_{\rm S},\rho]  
+ \mathcal{L}_{\rm{G}}\rho
+ \frac{P_c}{2} \mathcal{D}[X^-]\rho,
\end{equation}
where $P_c$ is an incoherent pump term,
with ${\cal D}[O] \rho =
2 O \rho O^\dagger - \rho O^\dagger O - O^\dagger O \rho$,
and  the cavity dissipator  is
\begin{align}\label{eq:GME_diss}
    \mathcal{L}_{\rm{G}}\rho &= \frac{1}{2} \sum\limits_{\omega,\omega'>0} 
    \Gamma_c(\omega) [ X^+(\omega)\rho  X^-(\omega') -  X^-(\omega') X^+(\omega)\rho] \nonumber \\ 
    &+
    \Gamma_c(\omega')[ X^+(\omega)\rho  X^-(\omega') - \rho X^-(\omega') X^+(\omega)].
\end{align}  
The dressed-state operators, $X^\pm$,
are defined from
\begin{equation}
     X^+(\omega)= \braket{j|\Pi_c|k} \ket{j}\bra{k},   
\end{equation}
where $\omega=\omega_k-\omega_j>0$, $ X^-=( X^+)^\dagger$, and $\Pi_c$ is a cavity operator linear in
the photon creation and destruction operators. We neglect
atom/emitter decay channels, since these are typically negligible. 
%
The cavity decay rates are obtained from 
$\Gamma_{c}(\omega) = 
    2 \pi J_{c} (\omega )$,
    where $J_{ c}(\omega)$ is the spectral bath function. 
    Below, 
we use $\Gamma_{c}=\kappa$, and show that this is sufficient to recover the classical spectral form if the appropriate
$\Pi_c$ operator can be identified.
    

\begin{figure}[b]
    \centering
\includegraphics[width=0.99\linewidth]{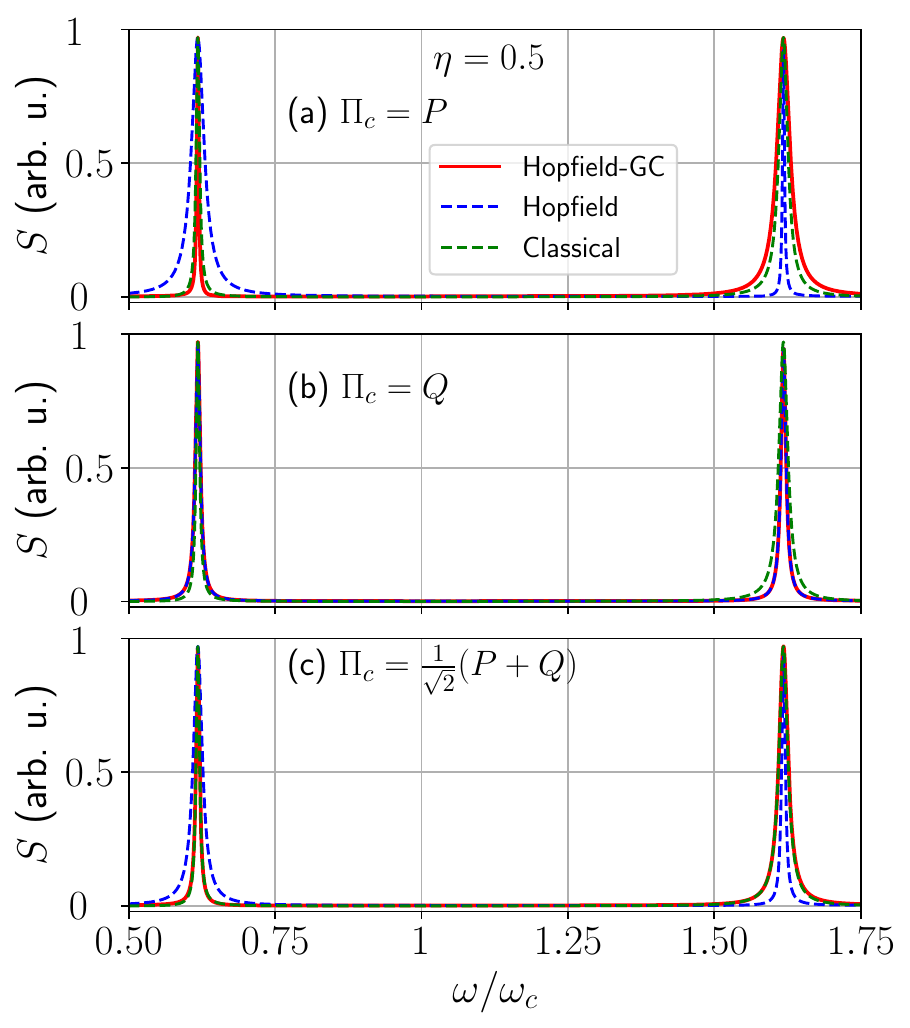}
\vspace{-0.5cm}
    \caption{ Hopfield GME
    results versus the classical model for
    $\eta=0.5$, and $\kappa=0.05g$, with three different bath models, (a) $\Pi_c=
    P$,
    (b) $\Pi_c
    =Q$,
    and (c) $\Pi_c
    =(P+Q)/\sqrt{2}$ ($\Pi_c = (P - Q)/\sqrt{2}$ gives identical results).
    Gauge-corrected results use
    $X^\pm_{\rm GC}$ and
    primed cavity operators for
    $\Pi_c$.
    Only model (c),
    with gauge corrections (`GC'), overlaps 
    with the classical solution.
    }
    \label{fig:spectra1}
\end{figure}

\begin{figure*}
    \centering
\includegraphics[width=0.95\linewidth]{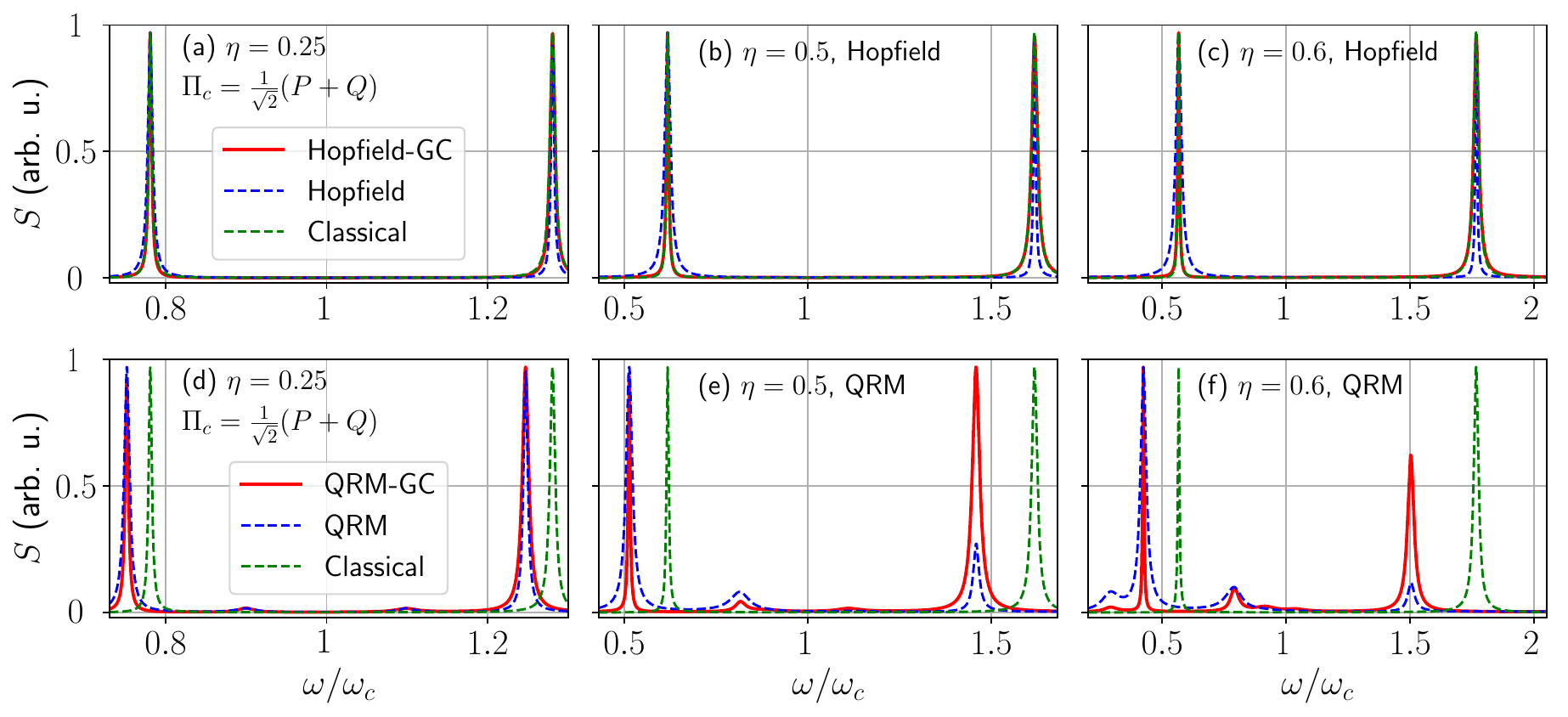}
\vspace{-0.3cm}
    \caption{Dissipative (GME) Hopfield model (a-c) and QRM (d-f) versus the classical solution, for three 
    values of $\eta$. We only present the  $\Pi_c
    =(P+Q)/\sqrt{2}$ bath model
    and show GME results, with and without gauge corrections. Once again, only the Hopfield model with gauge corrections overlaps with the classical solution. At higher values of $\eta$, the QRM also shows multiple resonances, and these are also substantially different  with gauge corrections. The QRM fails to recover the classical solution 
    in the USC regime. 
    }
    \label{fig:spectra2}
\end{figure*}

The precise 
form of 
$\Pi_c$ matters in the USC regime~\cite{Salmon2022}.
For example, one could choose
$\Pi_c=i(a^\dagger-a) \equiv P$,
or $\Pi_c=a^\dagger+a \equiv Q$, and obtain significantly different predictions, or any linear combination of these two.
This is not the case with a RWA.
Furthermore,
in the USC regime,  there a 
gauge ambiguity for the electric field operator~\cite{di_stefano_resolution_2019},
because  $P$
  represents the Coulomb gauge electric field, and we 
  are using  a system Hamiltonian in the dipole gauge. 
  For a restricted  TLS subspace, this ambiguity is corrected
  through the transformation $a'\rightarrow {\cal U} a {\cal U}^\dagger = a+i\eta\sigma_x$~\cite{Settineri2021Apr},
  where ${\cal U}=\exp(-i\eta(a+a^\dagger)\sigma_x)$ is the projected unitary operator~\cite{di_stefano_resolution_2019,savasta_gauge_2020,PhysRevA.107.013722}, with
   $\sigma_x= b+b^\dagger$
   (Hopfield model)
   or $\sigma_x=\sigma^++\sigma^-$ (QRM).
  Thus, one  must use
$a'$ and $a'^\dagger$ in the computation of the dissipators to ensure gauge invariance.
 The fact that the $\Pi_c$ operator should consist only of bosonic $a$ and $a^{\dagger}$ operators in the Coulomb gauge is a consequence of photon loss being associated only with electromagnetic degrees of freedom, and not the TLS~\cite{PhysRevA.107.013722}.

In the USC regime, the system has transition operators $\ket{j}\bra{k}$ which cause transitions between the dressed eigenstates of the system $\{\ket{j},\ket{k}\}$.
For the cavity mode operator, 
these transitions are obtained
from the dressed operators
$X^\pm$, and again these must be gauge corrected.
To make the notation clearer,
we can use
$X^\pm_{\rm GC}$ to indicate that we are applying
$\Pi_c$ operators with 
{\it gauge corrections}.
Thus, the  
cavity-emitted quantum spectrum is obtained from
\begin{equation}\label{eqn:spec}
    S^{\rm QM}\propto \text{Re} \left[\int_{0}^{\infty} d\tau e^{i \omega \tau} \int_{0}^{\infty} \left\langle X_{\rm GC}^-(t)X_{\rm GC}^+(t+\tau) \right\rangle dt\right],
\end{equation}
and calculations without gauge corrections simply use
$X^{\pm}$, i.e., without primed cavity operators in the computation of the dissipators
and cavity-mode obervables.
We will 
show both solutions to better highlight the role of these gauge corrections, and also show how they are {\it required} to recover classical correspondence.
In all GME calculations below, we use 
$P_c\ll \kappa$,
to ensure weak excitation,
and
the numerical results are carried out using
Python and QuTiP~\cite{johansson2012qutip,johansson_qutip_2013}.

It is important to stress that our gauge-corrected results are necessary to ensure gauge invariance.
For example, we could also use a Hopfield model in the Coulomb gauge, 
where
$
    H_{\rm Hop}^{\rm GC} = \omega_c a^\dagger a + \omega_0 b^\dagger b + ig\frac{\omega_0}{\omega_c}(b^{\dagger}-b)(a+a^\dagger)
    + D (a+a^\dagger)^2$,
and then we could use unprimed operators for the cavity mode, where indicated above, 
specifically in $X^\pm$
and $P$. Note also that $D$
must be the same in both gauges to ensure gauge invariance, which has been proven also for a dilute Dicke model~\cite{Garziano2020Aug}.
For the QRM, the gauge-corrected system Hamiltonian in the Coulomb gauge 
is~\cite{di_stefano_resolution_2019},
 $ {H}_{\rm QRM}^{\rm CG} =  \omega_c a^\dagger a
    + \frac{\omega_0}{2}
    \{ \sigma_z \cos(2 \eta (a+a^\dagger))+ \sigma_y \sin(2\eta (a + a^\dagger))\} $,
which  contains field operators to all orders.
For simplicity, we
use only the dipole gauge below,
but we have  checked that all results below are identical in the Coulomb gauge.

{\it Computed spectra.---}Figure~\ref{fig:spectra1},
shows the classical and quantum solutions for the dissipative Hopfield model,
 with three types of bath coupling 
models: (a) $\Pi_c=P$,
(b) $\Pi_c=Q$,
and (c)
$\Pi_c=(P\pm Q)/\sqrt{2}$, where primed indices are used 
for the  gauge-corrected models. 
We first choose
$\eta=0.5$ here, which is well into the USC regime,
with $\kappa=0.05g$.
As recognized,
{\it we find very good agreement with the classical solution
only when $1/\sqrt{2}(P\pm Q)$
and only with gauge corrections.}
To the best of our knowledge, this is the first time that such a solution and correspondence has been made,
and these results also demonstrate the significant problem with choosing an arbitrary system-bath interaction form in the USC regime. It should be noted that such a correspondence does not necessarily indicate that this is the \emph{correct} choice of dissipation model. Rather, this result allows for unambiguous connection and comparison between quantum and classical heuristic models of dissipation, and is further evidence of the limitations of purely phenomenological approaches, when treating losses in open quantum systems.

Next,
we look at the role of this
bath coupling model, for different $\eta$,
using $\Pi_c=(P\pm Q)/\sqrt{2}$,
 for both the Hopfield model and the QRM.
The spectral calculations are shown in 
 Fig.~\ref{fig:spectra2},
 along with the classical solution.
 We see that the Hopfield model and QRM differ substantially in all USC regimes, and the QRM takes on multiple resonances when $\eta$ is sufficiently large, even for weak excitation, as well as pronounced spectral asymmetries. Moreover, we find that the dissipative Hopfield model, with $\Pi_c=(P+Q)/\sqrt{2}$, agrees very well with the classical two oscillator model at all coupling regimes shown.
 This is clearly not the case for the 
 QRM, and such substantial differences should be easy to identify in experiments.

{\em Discussions and summary.}---We have shown how the optical spectra for a dissipative Hopfield model in USC can be described quantum mechanically or classically.
To achieve correspondence with 
the classical dissipative result, quantum models must properly respect gauge invariance and implement the appropriate bath coupling operator.
Without such a correspondence, 
any open-system master equations in this regime with {\it ad hoc} system-bath interactions are 
ambiguous
and can predict wildly differing spectra.

We have also clarified how the dissipative Hopfield model and QRM substantially differ, even for weak excitation, at all USC regimes, including the perturbative Bloch-Siegert regime. Thus, only the Hopfield model yields a classical correspondence under linear response, and this correspondence {\it only occurs with a careful treatment of quantum dissipation and gauge corrections}. 
While we 
used a normal mode expansion with heuristic broadening, this form is well established for high $Q$ cavities outside the USC regime, and future work could improve such models 
using 
classical and 
quantized quasinormal mode theories~\cite{PhysRevLett.122.213901,PhysRevResearch.2.033456}. 
%
One possible clue as to the significance of the $\Pi_c \propto P\pm Q$ coupling can be seen by noting that the classical phenomenological loss model is phase-insensitive. In the quantum loss model, a choice of $P \pm Q$ for the quadrature coupling to the bath is the only choice which gives equal coupling magnitude to each quadrature (i.e., is phase-insensitive in magnitude).

Broadly, these findings are important for a wide class of light-matter systems now emerging to study the USC regime, including lossy Landau systems and metallic systems~\cite{Zhang2016,Rajabali2021}.
Apart from showing a direct classical correspondence for dissipative modes,
our results can be used to guide 
open-system quantum models that are needed when observations are uniquely quantum in nature, e.g., in the QRM for any excitation including coherent excitation, and the Hopfield model excited with non-classical fields.

This work was supported by the Natural Sciences and Engineering
Research Council of Canada (NSERC), the National
Research Council of Canada (NRC), the Canadian Foundation
for Innovation (CFI), and Queen’s University, Canada. S.H.
acknowledges the Japan Society for the Promotion of Science
(JSPS) for funding support through an Invitational Fellowship.
F.N. is supported in part by Nippon Telegraph and Telephone
Corporation (NTT) Research, the Japan Science and
Technology Agency (JST) [via the Quantum Leap Flagship
Program (Q-LEAP), and the Moonshot R\&D Grant No. JPMJMS2061],
the Asian Office of Aerospace Research and Development
(AOARD) (via Grant No. FA2386-20-1-4069), and
the Foundational Questions Institute Fund (FQXi) via Grant
No. FQXi-IAF19-06.
We thank Jun Kono and Hideo Mabuchi for useful discussions.

\bibliography{references2}
\end{document}